\documentclass{elsart}

\usepackage{graphicx}

\usepackage{amssymb}

\newcommand{\Xjk}{X_{j,k}}
\newcommand{\Yjk}{Y_{j,k}}

\newcommand{\Xjplusk}{X_{j+1,k}}
\newcommand{\Yjplusk}{Y_{j+1,k}}
\newcommand{\xijplusk}{\xi_{j+1,k}}
\newcommand{\Omegajplusk}{\Omega_{j+1,k}}
\newcommand{\vjplus}{v_{j+1}}
\newcommand{\boldXjplusk}{\mathbf{X}_{j+1,k}}
\newcommand{\boldYjplusk}{\mathbf{Y}_{j+1,k}}

\begin{document}

\begin{frontmatter}


\title{Nonlinear modes of clarinet-like musical instruments}

\author{Daniel Noreland},
\author{Sergio Bellizzi},
\author{Christophe Vergez\corauthref{cor1}},
\corauth[cor1]{Corresponding author. Tel.: +33 491 16 41 63; fax: +33 491 16 40 12}
\ead{vergez@lma.cnrs-mrs.fr}
\ead[url]{http://www.lma.cnrs-mrs.fr/}
\author{Robert Bouc}
\address{Laboratoire de M\'{e}canique et d'Acoustique (LMA-CNRS, UPR7051),\\ 31 Chemin Joseph Aiguier, 13402 Marseille Cedex 20, France}

\begin{abstract}
The concept of nonlinear modes is applied in order to analyze the behavior of a model of woodwind reed instruments. Using a modal expansion of the impedance of the instrument, and by projecting the equation for the acoustic pressure on the normal modes of the air column, a system of second order ordinary differential equations is obtained. The equations are coupled through the nonlinear relation describing the volume flow of air through the reed channel in response to the pressure difference across the reed. The system is treated using an amplitude-phase formulation for nonlinear modes, where the frequency and damping functions, as well as the invariant manifolds in the phase space, are unknowns to be determined.
The formulation gives, without explicit integration of the underlying ordinary differential equation, access to the transient, the limit cycle, its period and stability. The process is illustrated for a model reduced to three normal modes of the air column.
\end{abstract}

\end{frontmatter}


\section{INTRODUCTION\label{sect-introduction}}
Musical wind instruments are interesting examples of nonlinear vibrating systems. The process governing the formation of
self-sustained oscillations is surprisingly complex.
In short, a wind instrument consists of a resonator and a generator~\cite{helmholtz77}. The resonator is the air column inside the instrument, and is usually characterized by the linear wave equation. The generator, in turn, consists of some kind of pressure or flow controlled valve~\cite{fletcher91}, where the relationship between air flow and pressure is starkly nonlinear.
Even a simplistic model of sound generation must include these nonlinear effects, simply because they are a prerequisite for the forming of a self-sustained oscillation from a continuous supply of air.
One important
reason why different wind instruments sound so different can be attributed to the nonlinearities, and how they interact with the vibrational modes of the air column at hand. The aim of this paper is to study how the sound production in a clarinet-like instrument can be analyzed and simulated in the framework of nonlinear modes. A first reason for pursuing this subject is to evaluate the possibility to use nonlinear modes to derive models of reduced complexity,
which are of interest for sound synthesis. Another goal is to identify important control parameters in a reduced model (functions of such entities as blowing pressure, pinching force and position of the player's lips on the reed etc) which can be regulated by a musician in an intuitive way, without a long period of training.

Modal analysis is the natural tool for characterizing linear
mechanical systems (in particular for numerical modelling, prediction
and experimental characterization). The extension of the modal theory
to nonlinear mechanical systems appears for the first time under the name
Nonlinear  Normal Modes   in the work  of Rosenberg~\cite{ros1962} for a
system of $n$-masses interconnected by nonlinear springs.  A NNM is defined as a  family of periodic
solutions of the equation of motions corresponding to
simple curves in  the configuration space.
In this paper
  the name Nonlinear Modes (NM) will be used in place of NNM.
Shaw and Pierre~\cite{sp1991} extended the concept of NM in the context
of phase space. This approach is
geometric by nature and makes use of the theory of invariant manifolds for
dynamical systems. A NM of an autonomous system is defined as a
two-dimensional invariant manifold in the phase space, passing "through a stable
equilibrium point of the system and, at that point, it is tangent to a
plan, which is an eigenspace of the system
linearized about  that equilibrium"~\cite{sp1993}. In the invariant manifold, the
modal dynamics reduces to a one-degree of freedom nonlinear
oscillator. This definition is valid for dissipative mechanical systems.
A nonlinear superposition technique is also proposed~\cite{sp1993} and its
validity is discussed by Pellicano and Mastroddi~\cite{pellicano97}.
The invariant manifold approach is also related to the methods based on
the theory of the normal forms~\cite{jl1991,nayfeh1993,touze2003}, where the invariant manifold and the modal
dynamics equation of motion are extracted from the minimal
representation. The construction of the NM for piecewise linear systems has been considered
By Jiang, Pierre and Shaw~\cite{spj2004}.
A review paper by Vakakis~\cite{va1997} and the book by Vakakis et al.~\cite{va1996} contain an
almost complete report of the history of the subject.

In this paper, a method devised by Bellizzi and Bouc~\cite{bb2006} for computing two-dimensional invariant manifolds of dynamical systems is considered.
This method (recalled in section \ref{sect-formulation}) can be run with systems with internal resonances without additional complexity, something that is often incompatible with other formulations~\cite{sp1991}.
Some comments are given in ~\cite{bb1999} regarding the access to higher-dimensional manifolds, but this approach
will not be considered here.
Indeed, resonance frequencies of wind instruments may be proportional to each others, in particular if the bore is a cylinder, which is the case for the clarinet  model,  detailed in section  \ref{s:basi_clari}.
The work presented in this paper should not be seen as a mere application of known methods to a particular example. The NM formulation employed here is very recent, and it is our hope that part of the presented work will contribute to the knowledge of how to compute the invariant manifolds (see more particularly sections \ref{sect-numsol} and \ref{ss:NM-compute}). In section \ref{s:NL_clari}, it is demonstrated how the computation of the  NM extending the linear modes of the bore, and of their properties, can be used to analyze and even predict the model's behavior in terms of transient and steady state, instantaneous amplitude and frequency, limit cycles and their stability. Results are systematically confronted with direct numerical simulations.

\section{THE CLARINET MODEL \label{s:basi_clari}}

A wide class of musical wind instruments have similar principles of functioning: the
player, by blowing inside the instrument destabilizes a valve (a
simple reed, a double reed or two lips). The acoustic response of the
instrument acts as a feedback loop which influences the valve
behavior. The production of a sound corresponds to the
self-sustained oscillation of this dynamical system. Obviously, in spite of
these similarities, the functioning of each class of instruments
possesses its own specificities. In this section, basic principles of the clarinet functioning are
briefly recalled. Simple models are available in the literature~\cite{backus63,nederveen69,worman71,wilson74,schumacher81}.

\begin{table}
\caption{Parameter values of the  clarinet model Eq.~(\ref{e:complete_model}) \label{tab:physparam}}
\begin{tabular}{llll}
entity&definition&value&unit\\
\hline
$\zeta$&embouchure parameter&0.35&1\\
$\gamma$&blowing pressure parameter&0.39&1\\
$\eta$&coefficient of viscosity&0.02&1\\
$Y_j$&admittance&$1.3\eta\sqrt{2j-1}$&1\\
$A$&$\zeta(3\gamma-1)/(2\sqrt\gamma)$&&1\\
$B$&$-\zeta(3\gamma+1)/(8\gamma^{3/2})$&&1\\
$C$&$-\zeta(\gamma+1)/(16\gamma^{5/2})$&&1\\
$c$&speed of sound&340&m/s\\
$l$&length of resonator&0.655&m\\
$\omega_j$&resonance frequency&$(2\mathrm{j}-1)2\pi c/(4l)$&1/s\\
\end{tabular}

\label{tab:params}
\end{table}

\subsection{The reed}

The reed is often modelled as a mass/spring/damper oscillator. However, because of a resonance frequency ($\simeq 2000$Hz) large compared to the first harmonics of typical playing frequencies, inertia and damping are often neglected~\cite{nederveen69}. This hypothesis leads, considering that reed dynamics is governed by the pressure difference across the reed, to  
\begin{equation}
k_s (z-z_0) = (p_{\mathrm{jet}}-p_{\mathrm{mouth}})
\label{clari_x}
\end{equation}
where $z$ (respectively $z_0$) is the reed position (respectively at rest). The reed is closed when $z=0$ and opened when $z > 0$. $k_s$ is the reed surface stiffness, $p_{\mathrm{mouth}}$ and $p_{\mathrm{jet}}$ are the pressure deviation in the mouth and under the reed tip, respectively.

\begin{figure}
\centerline{\includegraphics[width=7cm] {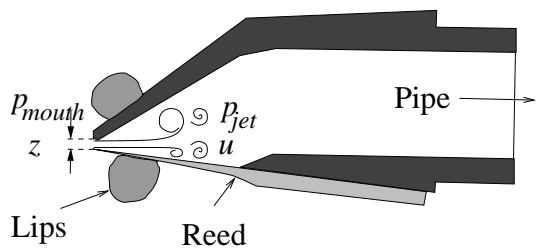}}
\caption{Scheme of the embouchure of a clarinet. \label{clari_cut}}
\end{figure}

\subsection{The air flow}
 As noted by Hirschberg~\cite{hirschberg94}, in the case of clarinet-like instruments, the control of the volume flow by the reed position is due to the existence of a turbulent jet. Indeed, a jet is supposed to form in the embouchure (pressure $p_{\mathrm{jet}}$) after the flow separation from the walls, at the end of the (very short) reed channel (see Fig. \ref{clari_cut}). Neglecting the velocity of air flow in the mouth compared to jet velocity $v_{\mathrm{jet}}$, the Bernoulli theorem applied between the mouth and the reed channel leads to
\begin{equation}
p_{\mathrm{mouth}} = p_{\mathrm{jet}} + \frac{1}{2} \rho v_{\mathrm{jet}}^2
\label{clari_u_first_form}
\end{equation}
where $\rho$ is the air density. Since the cross section at the inlet can be expressed as the product between the reed opening $z$ and the reed width $w_r$ (not visible in Fig. \ref{clari_cut} since it is transversal to the plane of the figure), Eq.~(\ref{clari_u_first_form}) can be re-written as
\begin{equation}
u= z w_r \sqrt{\frac{2}{\rho}(p_{\mathrm{mouth}}-p_{\mathrm{jet}})},
\label{clari_u}
\end{equation}
where $u$ is the volume flow across the reed. Combining Eqs. (\ref{clari_x}) and (\ref{clari_u}) leads to the well known expression of the volume flow as a function of the pressure difference across the reed  
\begin{equation}
u=  w_r (z_0 - \frac{1}{k_s}(p_{\mathrm{mouth}}-p_{\mathrm{jet}})) \sqrt{\frac{2}{\rho}(p_{\mathrm{mouth}}-p_{\mathrm{jet}})}.
\label{clari_xu}
\end{equation}

Since the cross section of the embouchure is large compared to the
 cross section of the reed channel, it can be supposed\cite[chapter7]{hirschberg95} that all the
 kinetic energy of the jet is dissipated through turbulence with no
 pressure recovery (like in the case of a free jet). Therefore, the
 pressure in the jet is (assuming pressure continuity) the acoustic
 pressure $p_r$ imposed by the resonator response to the incoming
 volume flow $u$.

Introducing the non-dimensional pressure~\cite[chapter 6]{kergomard95}
$\bar{p}=p_r/(k_s z_0)$ and volume flow $\bar{u}=Z_c u/(k_s z_0)$,
Eq.~(\ref{clari_xu}) reads
\begin{equation}
\bar{u} = \zeta (1+\bar{p}-\gamma)\sqrt{\gamma - \bar{p}}
\label{e:u_versus_p}
\end{equation}
with $\zeta = Z_c \omega_r  \sqrt{\frac{2 z_0}{k_s}}$ and $\gamma = p_{\mathrm{mouth}} / (k_s z_0)$. When the reed is closed, i.e. $1+\bar{p}-\gamma < 0$, the volume flow is zero ($\bar{u}=0$).

It has been verified~\cite{fritz2004a} that a cubic expansion of Eq.
(\ref{e:u_versus_p}) leads to a reasonably good approximation to the
resulting periodic solutions, at least far from the complete closing of
the reed. Therefore, we assume that the volume
flow is finally given by  
\begin{equation}
\bar{u} = u_0 + A \bar{p} +  B \bar{p}^2 + C \bar{p}^3
\label{e:u_versus_p_cubic}
\end{equation}
with $u_0=\zeta (1-\gamma)\sqrt{\gamma}$, $A=\zeta \frac{3\gamma
  -1}{2\sqrt{\gamma}}$, $B=-\zeta \frac{3\gamma
  +1}{8\gamma^{3/2}}$ and $C=-\zeta \frac{\gamma
  +1}{16\gamma^{5/2}}$.

\subsection{Acoustics in the instrument}


We consider a cylindrical bore (length $l$) for the clarinet, and follow
Debut\cite[page 60]{debut2004}.
Although the final model will be considered in the time domain, it is
first written in the frequency domain, where it is simplified before
going back to the time domain.

 The model retained is the wave equation inside the tube, with a
source at $x = x_s$ to take into account the air flow blown into the
instrument, and Neumann
and Dirichlet boundary conditions at the input and the
output of the tube, respectively:
\begin{equation}
\left\{
\begin{array}{lcc}
\displaystyle \left[\partial^2_{xx} - (\mathrm{j}\frac{\omega}{c} + \alpha)^2
\right] \bar{P}(\omega,x) & =
\displaystyle -\mathrm{j}\omega \frac{\rho}{S} U(\omega) \delta(x_s), & \forall x
\in [0,l],
\\
\displaystyle \partial_x \bar{P}(x, \omega) = 0  & &  \hbox{for } x = 0
\\
\displaystyle \bar{P}(x, \omega) = 0 & & \hbox{for } x = l.
\end{array}
\right.
\label{e:wave}
\end{equation}
$\alpha$ is a real number representing visco-thermal losses
(dispersion is neglected) and varies as the square root of the frequency.
The case of a clarinet model corresponds to a source located at the
input of the tube (i.e. $x_s \to 0$).

The dimensionless pressure field $\bar{P}(x,\omega)$ is decomposed into
the family  $\{f_{n}\}_{n\in \mathbb{N}}$  of
orthogonal eigenmodes of the air column inside the bore,
\begin{equation}
    \bar{P}(x,\omega)=\sum_{n=1}^{\infty}f_{n}(x) P_{n}(\omega).
\label{e:modal_p}
\end{equation}
In the case of a closed/open
cylindrical bore of length $l$ and dispersion neglected,
$f_{n}(x)= \cos{k_{n}}(x)$,  where $k_{n}=\frac{n\pi}{2l}$ and $n$
is an odd positive integer.

Modal coordinates $P_{n}(\omega)$ are calculated through the projection of
Eq.~(\ref{e:wave})  (with $\bar{P}$ replaced by Eq.
(\ref{e:modal_p}), truncated to $N$ modes) on each mode $f_{n}$, leading to
\begin{equation}\label{eq:systneuua}
 -\omega^2 P_n(\omega)+2\alpha c\ \mathrm{j} \omega
P_n(\omega)+(\omega_n^2-\alpha^2 c^2)\ P_n(\omega)=\frac{2c}{l} \mathrm{j} \omega
U(\omega)
\end{equation}
where $\omega_n = k_n c$. Several approximations are now made concerning
the losses coefficient $\alpha$.
Since the damping of each mode is small\cite[page 70]{debut2004} ($\alpha
c << \omega_n$), the
third term of the left hand side of Eq.~(\ref{eq:systneuua}) can be
reduced.
Moreover, though $\alpha$ is a function of the frequency, we consider
that a constant value $\alpha_n$ can be associated to each mode
(\cite[page 72]{debut2004}). Noting $\alpha_n \simeq \frac{Y_n}{l}$,
where $Y_n$ is the value of the admittance at frequency $\omega_n/2
\pi$, Eq.
(\ref{eq:systneuua}) can now be written in the time domain as a second
order ODE:
\begin{equation}\label{eq:systnequa_2}
 \ddot{p}_n(t)+2Y_n\frac{c}{l} \dot{p}_n(t)+\omega_n^2
p_n(t)=\frac{2c}{l}\dot{\bar{u}}(t),
\end{equation}
where $p_n(t)$ is the inverse Fourier transform of $P_n(\omega)$.

\subsection{Complete model}
Considering Eqs. (\ref{eq:systnequa_2}) and
 (\ref{e:u_versus_p_cubic}) leads to the dimensionless model, made of
 $N$ second-order ODE,
\begin{equation}\label{e:complete_model}
\begin{array}{l}
 \ddot{p}_n(t)+2Y_n\frac{c}{l} \dot{p}_n(t)+\omega_n^2
 p_n(t)=\\
 \displaystyle \frac{2c}{l} \left( A  + 2B \sum_{i=1}^N p_i (t) + 3C \left(
 \sum_{i=1}^N p_i (t) \right)^2 \right ) \sum_{i=1}^N \dot{p}_i (t).
\end{array}
\end{equation}
The total pressure at the input end of the instrument is given by the sum of all $p_n(t)$ (Eq.~(\ref{e:modal_p}) with $x=0$).

The intention now is to apply the concept of NM to this system of Eqs. (\ref{e:complete_model}). The aim is to get a reduced representation of the model, which could help in analyzing the model behavior, and which could be useful for sound synthesis purpose.

\section{NONLINEAR MODES}

\label{sect-formulation}

It is briefly recalled here how to characterize the NM in the framework of the
invariant manifold theory using an amplitude-phase transformation according to Bellizzi and Bouc
~\cite{bb2005,bb2006}.

We consider a system of the form
\begin{equation}
\mathbf{M}\ddot{\mathbf{P}}(t)+\mathbf{F}(\dot{\mathbf{P}}(t),\mathbf{P}(t))=\mathbf{0}\label{eq1}
\end{equation}
where $\mathbf{P}$ is an $n$-vector function, $\mathbf{M}$
is a non-singular symmetric square $N\times N$-matrix and $\mathbf{F}$
is a (sufficiently regular) vector function with dimension $N$ such
that $\mathbf{F}(\mathbf{0},\mathbf{0})=\mathbf{0}$.

\subsection{Linear modes as a starting point \label{sect-linearcase}}
In this section, definition and properties of the normal modes are recalled when $F$ in Eq.~(\ref{eq1}) is a linear function. Variables introduced in this section will be then extended to the nonlinear case in further sections.

\subsubsection{Undamped case}
Let us consider the case where $\mathbf{F}(\mathbf{P}, \mathbf{\dot{P}})=\mathbf{K}\mathbf{P}$, where $\mathbf{K}$ is a symmetric square $N \times N-$matrix. The normal modes are then the $N$ pairs $(\Omega_p, \mathbf{\Psi_p})$ solutions of the eigenvalue problem
\begin{equation}
\mathbf{K} \mathbf{\Psi_p} = \mathbf{M} \mathbf{\Psi_p} \Omega_p^2
\end{equation}
and  the orthogonality condition and the mass-normalization are written respectively as
\begin{eqnarray}
 \mathbf{\Psi_p^t}  \mathbf{M} \mathbf{\Psi_q} &  = 0 & \qquad \forall p \neq q,  \\
 \mathbf{\Psi_p^t}  \mathbf{M} \mathbf{\Psi_p} &  = 1 & \qquad \forall 1 \leq p \leq n.
\end{eqnarray}

A family of periodic solutions is associated to each normal mode as
\begin{equation}
\mathbf{P(t)} = v \mathbf{X}(\phi(t))
\end{equation}
where
\begin{equation}
\left\{ \begin{array}{lcl}
\mathbf{X}(\phi) & = & \mathbf{\Psi_p} \cos (\phi) \\
\phi(t) & = & \Omega_p t + \varphi \\
v & = & a   \; \mbox{(constant)}
 \end{array}\right. . \end{equation}

The amplitude and frequency of the periodic motion are $v$ and $\Omega_p/2\pi$ respectively and $\mathbf{X}$
is a periodic function with respect to the phase variable $\phi$.

\subsubsection{Damped case}

 If $\mathbf{F}(\mathbf{P},\dot{\mathbf{P}})=\mathbf{K}\mathbf{P}+\mathbf{C}\dot{\mathbf{P}}$,
where $\mathbf{K}$ and $\mathbf{C}$ are real,
square matrices, the eigenvalue problem to solve is now
~\cite{meirovitch1980}
%
%
%
\[
\left(\begin{array}{cc}
{\mathbf{C}}& {\mathbf{M}}\\
{\mathbf{M}} & {\mathbf{0}}\end{array}\right){\mathbf{\Psi}_p^d}\lambda_p+\left(\begin{array}{cc}
{\mathbf{K}} & {\mathbf{0}}\\
{\mathbf{0}} & -{\mathbf{M}}\end{array}\right){\mathbf{\Psi}_p^d}=0\]
 with $\lambda_p=\eta_p \pm \mathrm{j}\Omega_p$ (assuming $\Omega\neq0$) and
 ${\mathbf{\Psi}_p^d}=\left({\mathbf{\Psi}_p}^{T},\lambda{\mathbf{\Psi}_p}^{T}\right)^{T}$.
 With  ${\mathbf{\Psi}_p}= \mathbf{\Psi}_p^c+\mathrm{j} \mathbf{\Psi}_p^s$,
the orthonormalization
 condition can be selected as
 \[
\begin{array}{c}
\mathbf{\Psi}_p^c{^{T}}\mathbf{M} \mathbf{\Psi}_p^c+ \mathbf{\Psi}_p^s{^{T}}\mathbf{M} \mathbf{\Psi}_p^s=1,\\
\mathbf{\Psi}_p^c{^{T}}\mathbf{M}\mathbf{\Psi}_p^s=0.\end{array}\]

The family of solutions associated to each mode is now written as
\begin{equation}
\mathbf{P(t)} = v(t) \mathbf{X}(\phi(t))
\end{equation}
where
\begin{equation}
\left\{ \begin{array}{lcl}
\mathbf{X}(\phi) & = & \mathbf{\Psi}_p^c\cos\phi-\mathbf{\Psi}_p^s\sin\phi\\
\phi(t) & = & \Omega_p t + \varphi \\
v(t) &=& a e^{\eta_p t} \\
 \end{array}\right. . \label{eqlc1}\end{equation}

Depending on the sign of $\eta_p$, the motion can be damped ($\eta_p<0$), amplified ($\eta_p>0$), or periodic ($\eta_p=0$).
The amplitude and frequency of the motion are $v(t)$ and $\Omega_p/2\pi$ respectively. Here also $\mathbf{X}$
is a periodic function with respect to the phase variable $\phi$.

\subsection{Definition of nonlinear modes}

In the case of a nonlinear function $\mathbf{F}$ in Eq.~(\ref{eq1}), the linear modal formalism recalled in section \ref{sect-linearcase} is extended hereafter with the following major differences:
\begin{itemize}
\item $\mathbf{X}$ is not only a periodic function with respect to the phase variable but is also a function of the amplitude $v$ (see Eq.~(\ref{eq2})).
\item Even in the case of a periodic motion, the amplitude  $v$ is a function of time defined through a differential equation  (see Eq.~(\ref{eq3}a). The linear damped case studied in the above section would correspond to $\xi(t)=\eta$ in Eq.~(\ref{eq3}a).
\item The pulsation of the motion $\Omega$ is no more constant, but is a function of the amplitude $v$ and the phase $\phi$ (see Eq.~(\ref{eq3}b).
\end{itemize}
Indeed, we focus on motions (solutions of Eq.~(\ref{eq1})) where the pressure components and
its derivatives ($\mathbf{P}$ and $\dot{\mathbf{P}}$)
are related to a single pair of amplitude and phase variables ($v$ and $\phi$)
according to
\begin{equation}
\left\{ \begin{array}{lcl}
\mathbf{P}(t) & = & v(t)\mathbf{X}(v(t),\phi(t))\\
\dot{\mathbf{P}}(t) & = & v(t)\mathbf{Y}(v(t),\phi(t))\end{array}\right.\label{eq2}\end{equation}
where $\mathbf{X}$ and  $\mathbf{Y}$ are $N$-vector functions,
which are {\it $2\pi$-periodic with respect to the phase variable $\phi$} and the amplitude and phase
variables are governed by the two first-order differential equations
\begin{equation}
\left\{ \begin{array}{lcl}
\dot{v}(t) & = & v(t)\xi(v(t),\phi(t))\;\\
\dot{\phi}(t) & = & \Omega(v(t),\phi(t))
\;\end{array}\right.\; \mbox{with}\;
\left\{ \begin{array}{lcl}
v(0)&=& a\\
\phi(0) &=& \varphi
\end{array}\right. .\label{eq3}\end{equation}
In Eq.~(\ref{eq3}), $\Omega$ (the frequency function or frequency modulation function)
and $\xi$ (the damping function or amplitude modulation function) are two scalar functions.
As established in \cite{bb2006}, these two functions can be only chosen as {\it even and
$\pi$-periodic with respect to the phase variable}. Furhermore, $\varphi$
$\in[0,2\pi]$ and $a>0$ are two given constants which set
the initial conditions of the motion.

If such a family of motions (\ref{eq2})--(\ref{eq3}), parameterized by the variables $(a,\varphi)$ exist, it defines a NM for Eq.~(\ref{eq1}), which is characterized
by the four functions $\mathbf{X}$, $\mathbf{Y}$, $\Omega$ and $\xi$.

\subsection{Some properties of nonlinear modes} \label{NLMprop}
For a given NM, the modal motions are confined to lie on a
 two-dimensional invariant
manifold~\cite{sp1991,ac2006} in the phase space, defined by the parametric equations
\begin{equation}
\left\{ \begin{array}{lcl}
\mathbf{P} & = & a\mathbf{X}(a,\varphi)\\
\dot{\mathbf{P}} & = & a\mathbf{Y}(a,\varphi)\end{array}\right. \; \mbox{for} \; (a,\varphi) \in {\mathbf{R}}\times[0,2\pi]. \label{eq2c}
\end{equation}
and the modal dynamics  on the invariant manifold are given by
Eq.~(\ref{eq3}). In terms of signal processing, the function $\dot{\phi}$ characterizes the instantaneous frequency of the modal motion and the function $v$ defines the amplitude modulation of the modal motion.

If the damping function $\xi\equiv0$ (which means that $v(t)=a, \; \forall t$), all the modal motions
(defined by Eqs. (\ref{eq2})
and (\ref{eq3})) will be periodic. The period is given by
\begin{equation}
T(a)=\int_{0}^{2\pi}\frac{1}{\Omega(a,\phi)}d\phi,\label{eq2d}
\end{equation}
showing that the period is only amplitude dependent. This situation appears for autonomous conservative systems~\cite{bb2005}.

Periodic modal motions may also exist if $\xi \not\equiv0$ or more precisely if $\frac{\xi}{\Omega} (\not\equiv0)$ does not keep
a constant sign. Indeed, from Eq.(\ref{eq3}), it follows
that\begin{equation}
\frac{dv}{d\phi}=v\tau(v,\phi)\label{eq1ddl16}\end{equation}
where $\tau(v,\phi)=\displaystyle\frac{\xi(v,\phi)}{\Omega(v,\phi)}$
can be viewed as a "generalized damping rate
function".
Since $\tau$ is $\pi$-periodic
with respect to the independent variable $\phi$, a periodic solution $v^*$ (with $v^*(\phi)=v^*(\phi+\pi)$)
 may exist for some $\xi$ and $\Omega$ (one necessary condition
being that $\tau(v,\phi)$ does not keep a constant sign). It follows
that the associated modal motion \begin{equation}
\mathbf{P}(t)=v^*(\phi(t))\mathbf{X}(v^*(\phi(t)),\phi(t))\label{eq1ddl16bb}
\end{equation}
with
$\dot{\phi}(t)  =  \Omega(v^*(\phi(t)),\phi(t)) \; \mbox{and}
\;  
\phi(0) = \varphi$,
 will be $T$-periodic with period \begin{equation}
T=\int_{0}^{2\pi}\frac{d\phi}{\Omega(v^{*}(\phi),\phi)}.\label{eq1ddl16b}
\end{equation}
From Eq.~(\ref{eq1ddl16}), the stability analysis of the periodic function $v^*$ can be deduced
using the average principle in the context of perturbation theory~\cite{hale69}
from the existence of an
equilibrium point in the averaged equation
\begin{equation}
\frac{dv}{d\phi}=v <\tau>(v) \label{eq1ddl16m}\end{equation}
where
\begin{equation} <\tau>(v)=\frac{1}{2\pi}\int_0^{2\pi}\frac{\xi(v,\phi)}{\Omega(v,\phi)}\mathrm{d}\phi. \label{eq1ddl16mm}
\end{equation}

More precisely, each equilibrium point $v^{**}$ (defined as $<\tau>(v^{**})=0$), which can be viewed as  a constant approximation of a periodic function $v^*$, characterizes a periodic modal motion (or limit cycle)
on the invariant manifold. This limit cycle is asymptotically stable if  $\frac{d<\tau>}{dv}(v^{**}) < 0$.
Note that in order to analyze the stability in the complete phase space and not only in the invariant manifold, the Floquet theory has to be applied~\cite{hale69,bb2006}.
Finally, the periodic modal motion and the associated period can be approximated by, respectively,
\begin{equation}
\mathbf{P}(t)=v^{**}\mathbf{X}(v^{**},\phi^{**}(t))\label{eq1ddl16bbb}\end{equation}
with
$\dot{\phi}^{**}(t)  =  \Omega(v^{**},\phi^{**}(t)) \; \mbox{and}
\;  
\phi^{**}(0) = \varphi$, and
\begin{equation}
T=\int_{0}^{2\pi}\frac{d\phi}{\Omega(v^{**},\phi)}.\label{eq1ddl16bbbb}\end{equation}

\subsection{Characterization of a nonlinear mode}
Substituting Eq.~(\ref{eq2}) into Eq.~(\ref{eq1}) and using Eq.
(\ref{eq3}) yields a set of first-order nonlinear
Partial Differential Equations (PDEs) in the two variables $(v,\phi)$,
\begin{equation}
\left(\mathbf{X}+v\mathbf{X}_{v}\right)\xi+\mathbf{\mathbf{X}_{\phi}}\Omega=\mathbf{Y},\label{eq4a}\end{equation}
\begin{equation}
\mathbf{M}\left(\mathbf{Y}+v\mathbf{Y}_{v}\right)\xi+\mathbf{M}\mathbf{Y}_{\phi}\Omega+\frac{1}{v}\mathbf{F}(v\mathbf{Y},v\mathbf{X})=\mathbf{0}\label{eq4b}\end{equation}
where $(.)_{\phi}$ and $(.)_{v}$ denote the partial differentiation with respect to $\phi$ and $v$, respectively.
The PDEs (\ref{eq4a})--(\ref{eq4b}) are independent of time.

In order to characterize the four unknown functions (of $v$ and $\phi$) $\mathbf{X}$, $\mathbf{Y}$,
$\Omega$ and $\xi$, it is necessary to add two scalar constraint
equations to (\ref{eq4a}) and (\ref{eq4b}) (often called normalization
conditions). Due to the $2\pi$-periodicity with respect to the variable
$\phi$, the functions $\mathbf{X}$  can be expressed
as \begin{equation}
\nonumber
 \begin{array}{lcl}
\mathbf{X} & = & \mathbf{X}^{oc}+\mathbf{X}^{ec}+\mathbf{X}^{os}+\mathbf{X}^{es}
\end{array}
\label{eq5}\end{equation}
where $(.)^{oc}$, $(.)^{ec}$, $(.)^{os}$, and $(.)^{es}$
denote the odd cosine, even cosine, odd sine, and even sine
terms in the corresponding Fourier expansions. We will adopt in this study, without loss
of generality, the following scalar constraint equations
\begin{equation}
\mathbf{X}^{oc^{T}}\mathbf{M}\mathbf{X}^{oc}+\mathbf{X_{\phi}}^{os^{T}}\mathbf{M}\mathbf{X_{\phi}}^{os}=\cos^{2}\phi,
\label{eq6}
\end{equation}
\begin{equation}
\mathbf{X}^{oc^{T}}\mathbf{M}\mathbf{X_{\phi}}^{os}=0.\label{eq7}\end{equation}
These constraint equations involve only odd terms of the sine/cosine developments due to the
assumptions imposed on the two scalar functions $\Omega$ and $\xi$ (see Eq. (\ref{eq3})). Moreover,
this choice  reduces to usual normalization conditions
for some important special cases~\cite{bb2005} including  the linear case (see next section).

Finally, a NM of the system (\ref{eq1}) is obtained by solving
Eqs. (\ref{eq4a})--(\ref{eq7}) for the four
functions $\mathbf{X}$, $\mathbf{Y}$, $\Omega$ and $\xi$ with initial values given at $v=0$ and the periodicity
properties
\begin{equation}\label{omegaksiper}
\begin{array}{l}
\mathbf{X}(v,\phi)=\mathbf{X}(v,\phi+2\pi),\\
\Omega(v,\phi)=\Omega(v,-\phi)=\Omega(v,\phi+\pi),\\
\xi(v,\phi)=\xi(v,-\phi)=\xi(v,\phi+\pi).
\end{array}
\end{equation}

It can be shown~\cite{bb2005} that a NM can be defined from each mode of
the underlying linear system by selecting it as an initial condition, using the relations (\ref{eqlc1}).
However, contrary to the case for linear systems, an $N$ DOF nonlinear mechanical system can possess more that $N$
NM~\cite{va1997}.

It is worth noting that depending on the properties of the function $\mathbf{F}$,
some functions among $\mathbf{X}^{os}$, $\mathbf{X}^{es}$, $\mathbf{X}^{oc}$,
$\mathbf{X}^{ec}$ can be discarded.
For example,
\[
\begin{array}{lll}
\mbox{if $\mathbf{F}(\mathbf{P},\dot{\mathbf{P}})=\mathbf{F}(\mathbf{P})$} & \mbox{then} & \begin{array}{ll}
\mathbf{X}^{os}\equiv\mathbf{0}, & \mathbf{X}^{es}\equiv\mathbf{0}
\end{array},\\
\mbox{if $\mathbf{F}(\mathbf{P},\dot{\mathbf{P}})=-\mathbf{F}(-\mathbf{P},-\dot{\mathbf{P}})$} & \mbox{then} &
\begin{array}{ll}
\mathbf{X}^{ec}\equiv\mathbf{0}, & \mathbf{X}^{es}\equiv\mathbf{0}.
\end{array}
\end{array}
\]

%


\subsection{Numerical solution of the equations describing the manifold}
\label{sect-numsol}
Eqs. (\ref{eq4a})--(\ref{eq7}) constitute a partial differential algebraic equation (PDAE). It is an initial-boundary-value problem,
where $v$ acts as a time-like variable, and with periodic boundary conditions in the $\phi$--direction.
Differential algebraic equations are generally much more difficult to solve than differential equations. Firstly, the initial conditions must satisfy not only the algebraic constraints, but also a number of compatibility equations depending on the index of the equation~\cite{hairwan96}.
For the solutions considered in this work, the NM are the continuations of corresponding linear modes. Consistent initial conditions of the PDAE are therefore directly obtained from the corresponding linearized system, or equivalently by setting $v=0$ in Eqs. (\ref{eq4a})--(\ref{eq7}), which then collapse to an algebraic system of equations. Secondly, depending on the formulation of the numerical method, there is often a numerical drift in the fulfilment of the algebraic constraints as the integration advances.
 

The PDAE is solved by the finite difference method. The unknowns are discretized in the $\phi$--direction, after which the occurring derivatives with respect to $\phi$ are approximated using finite difference approximations.
This so-called \emph{method of lines}~\cite{GKO95}, where the $v$-derivatives are still left on their continuous form, leaves us with a usual differential algebraic equation (DAE), which can be solved with a suitable numerical integration scheme in the $v$--direction. The \emph{backward differentiation formulae} (BDF) are a wide class of methods for DAEs, of which the implicit Euler scheme is the most well-known representative.
The approach differs from those used by Pesheck et al.~\cite{psp2002} or Bellizzi and Bouc~\cite{bb2006}, where use is made of Galerkin methods based on trigonometric terms, and in the latter case polynomial terms in the $v$-direction. Although elegant, the Galerkin treatment becomes prohibitively complex as the number of expansion terms increases. A low order, implicit scheme for step-wise advancement in the $v$-direction
is also
better adapted for capturing variations, or even irregularities, in the $v$-direction of the solution, which do not readily lend themselves to an accurate description with a polynomial basis. A step-wise integration is also consistent with the initial value character of the equation in the sense that at each $v$, the solution depends on the earlier, but not on subsequent states.

Let
$Z(v,\phi)$ be anyone of the unknowns $\xi(v,\phi)$, $\Omega(v,\phi)$, $X^i(v,\phi)$,
$Y^i(v,\phi)$ for $i=1,\ldots,n$ and $\mathbf{X}(v,\phi)=(X^1(v,\phi),X^2(v,\phi),\ldots,X^n(v,\phi))^T$.
The approximation $U$ is defined through the discretization
\begin{eqnarray*}
Z(v_j,\phi_k) \approx U_{j,k},\;\; & v_j=jh_v, & j=0,1,\ldots, \\
{}&\phi_k=kh_\phi,\;\; & k=0,1,...,N_\phi-1, \;\; h_\phi=2\pi/N_\phi.
\end{eqnarray*}
In the sequel, the respective discrete approximations $U_{j,k}$ are denoted $X_{j,k}^i$, $Y_{j,k}^i$ etc.
where the meaning is clear.
Since the problem is periodic in the $\phi$--direction, and can be expected to have a smooth solution in this direction,
it is natural to use a \emph{pseudo-spectral}~\cite{gotors77} approximation of the $\partial/\partial\phi$-terms.
The approximation can be interpreted as a usual finite difference approximation where the number of points, and hence
the order of accuracy, is a function of the step-size $h_\phi$ in the direction of differentiation.
The implementation involves manipulations in the Fourier space and relies on the fast Fourier transform applied to the grid data.
Accordingly, a term $\partial Z(v_k,\phi_k)/\partial \phi$ is
approximated by the difference scheme
\[
\frac{\partial Z(v_j,\phi_k)}{\partial \phi} \approx D_\phi U_{j,k}  
\triangleq \sum_{m=-N_\phi/2}^{N_\phi/2} d_m U_{j,\mbox{mod}(k+m,N_\phi)},
\]
where $D_\phi$ is denoted the pseudo-spectral differential operator, and the coefficients $d_m$ depend on the discretization.
The second index of $U$ reflects the fact that data from a grid---periodic in the $\phi$-direction---is used in a wrap-around fashion.
For the appropriate choice of $d_m$, the approximation has  
spectral convergence rate, meaning that the error decreases faster than any polynomial in $h_\phi$.

Leaving the system of equations (\ref{eq4a})--(\ref{eq7}) stated on its implicit form, and
by approximating the appearing derivatives with the
pseudo-spectral scheme in the $\phi$-direction and a backward difference in the $v$-direction,
the implicit Euler approximation of the PDAE at hand is
given by

\begin{equation}
\begin{array}{l}
\Xjplusk^i+\vjplus\frac{\Xjplusk^i-\Xjk^i}{h_v}\xijplusk + D_\phi \Xjplusk^i \Omegajplusk = \Yjplusk^i, \label{euler1}  \\
i=1,\ldots,n, \; k=0,\ldots,N_\phi-1  \\
\sum_{l=1}^n M_{il} \left(\Yjplusk^l + \vjplus\frac{\Yjplusk^l-\Yjk^l}{h_v}\xijplusk  +
D_\phi\Yjplusk^l\Omegajplusk \right) +
\frac{F_i(\vjplus\boldXjplusk,\vjplus\boldYjplusk)}{\vjplus}=0 \label{euler2} \\  
 i=1,\ldots,n, \; k=0,\ldots,N_\phi-1  \\
\boldXjplusk^{oc^T}[\mathbf{M}]\boldXjplusk^{oc}+D_\phi\boldXjplusk^{os^T} [\mathbf{M}]
D_\phi\boldXjplusk^{os} = \cos^2\phi_k, \;\;\; k\in \{\mathrm{LI}\}\\
\boldXjplusk^{oc^T}[\mathbf{M}]D_\phi\boldXjplusk^{os}=0, \;\;\; k\in \{\mathrm{LI}\} .
\end{array}
\end{equation}

The decomposition in even and odd cosine and sine parts is done with the aid of the discrete Fourier transform (DFT),
followed by a selection of the appropriate Fourier components, and finally an inverse DFT.  
Due to the symmetric properties of the unknowns, not all the equations $k=0,\ldots,N_{\phi}-1$ are linearly
independent.
This reduces the number of equations, but since it is known
beforehand
that $\Omega$ and $\xi$ can
be expanded in only even cosine terms, it is possible to reduce the number of unknowns correspondingly.
Accordingly, the set $\{\mathrm{LI}\}$ is chosen so that only linearly independent equations are retained,
and new variables $\xi^{ec}$ and $\Omega^{ec}$, containing only the odd cosine components of $\xi$ and $\Omega$,
are introduced as unknowns in the numerical solution of Eq.~(\ref{euler1}).

The solution on each new $v$-level $j+1$ is obtained by solving
Eq.~(\ref{euler1}) for all $\Xjplusk^i$, $\Yjplusk^i$, $\xijplusk$ and $\Omegajplusk$. This non-linear system of equations is solved numerically using the Newton method, where the starting solution in each step is obtained as a first order extrapolation of the solution at
levels $j-1$ and $j$.

\subsection{Computation of time dependent solutions from the manifolds}
\label{ss:NM-compute}
Once the surfaces $\mathbf{X}(v,\phi)$, $\mathbf{Y}(v,\phi)$, $\xi(v,\phi)$ and $\Omega(v,\phi)$ have been computed, the time evolutions $v(t)$ and $\phi(t)$ can be computed numerically solving Eq.~(\ref{eq3}). Since numerical approximations of $\xi$ and $\Omega$ are known only for certain discrete values $v_i$ and $\phi_i$, a two-dimensional interpolation procedure is used, employing trigonometric interpolation in the $\phi$-direction and quadratic interpolation in the $v$-direction. For solutions representing a limit cycle, it is necessary to solve Eqs. (\ref{eq4a})--(\ref{eq7}) on an interval $[0,v_{\mathrm{max}}]$, where $v_{\mathrm{max}}$ is large enough so that the computed region of the invariant manifold  contains the limit cycle. In practice, this implies a value of $v_{\mathrm{max}}$ slightly above the amplitude of the limit cycle $v^{**}$ as defined in section \ref{NLMprop}. An estimate of when $v_{\mathrm{max}}$ is reached can be obtained by keeping track of the \emph{mean damping function}
Eq.~(\ref{eq1ddl16mm}),
which is a measure of the average energy dissipated or supplied to the system over one period. It is zero at $v=v^{**}$.
The physical variables in phase space are finally given by Eq.~(\ref{eq2}), where once again the two-dimensional interpolation procedure is employed.

\section{Nonlinear modes for the clarinet model}
\label{s:NL_clari}

\begin{figure}
\centerline{\includegraphics[width=0.9\columnwidth]{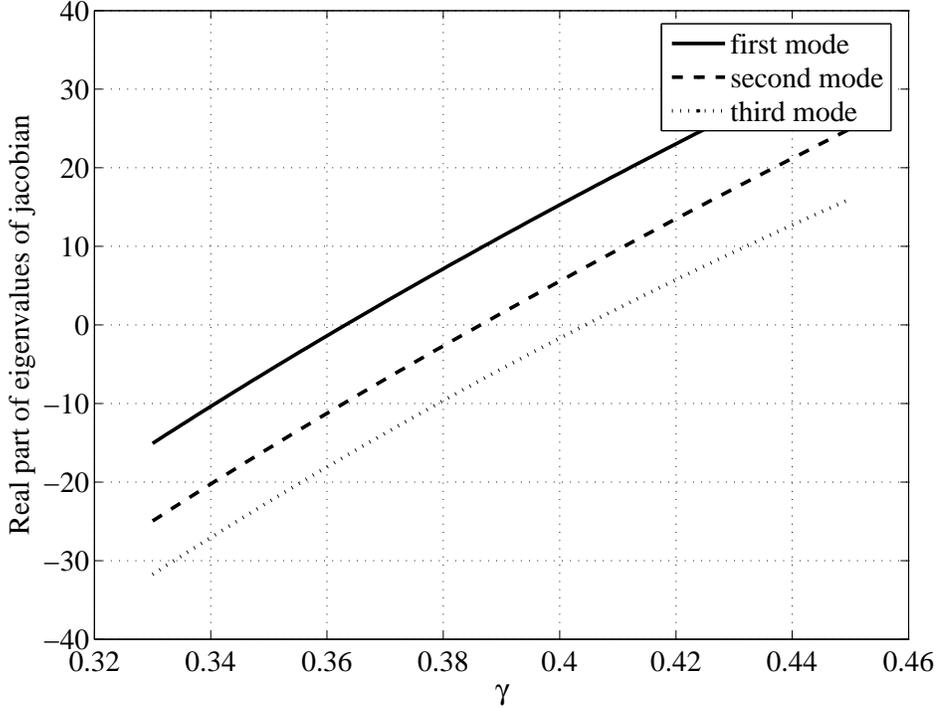} }
\caption{Real parts of the eigenvalues of the system matrix of a linearized, first order version of Eq.~(\ref{e:complete_model})  as a function of $\gamma$ (with parameter values given in Tab.~\ref{tab:params}). \label{eigenvalues}}
\end{figure}

The clarinet model described in section \ref{s:basi_clari} is considered for a case where $N=3$. As the excitation of modes four and onwards is fairly weak for the chosen blowing pressure, only the most prominent three modes are treated for clarity.
The method presented in section \ref{sect-formulation} is applied to find the NM of the system, with
$\mathbf{P}=[p_1, p_2, p_3]^T$. 
With Eq.~(\ref{e:complete_model}) linearized and written on first-order form, it is possible to determine the value of the blowing parameter $\gamma$ where an equilibrium point loses its stability and a self-sustained oscillation can appear if a Hopf-bifurcation occurs. The motion becomes oscillatory when the corresponding eigenvalue of the linearized system matrix crosses the imaginary axis.
The model parameters (Tab.~\ref{tab:params}) are chosen so as to correspond to a \emph{mezzo forte} playing condition. The first vibrational mode becomes linearly unstable at $\gamma=0.363$, see Fig.~\ref{eigenvalues}. For $\gamma=0.386$, also the second mode becomes linearly unstable. Thus, the chosen blowing pressure ($\gamma=0.39$) is just strong enough to render also the second mode linearly unstable. It means that for the chosen value of $\gamma$, playing in the first or second register might be possible, depending on their stability and on the initial conditions. This will be investigated in the following.

\subsection{First mode}
\begin{figure}
\centerline{\includegraphics[width=0.9\columnwidth]{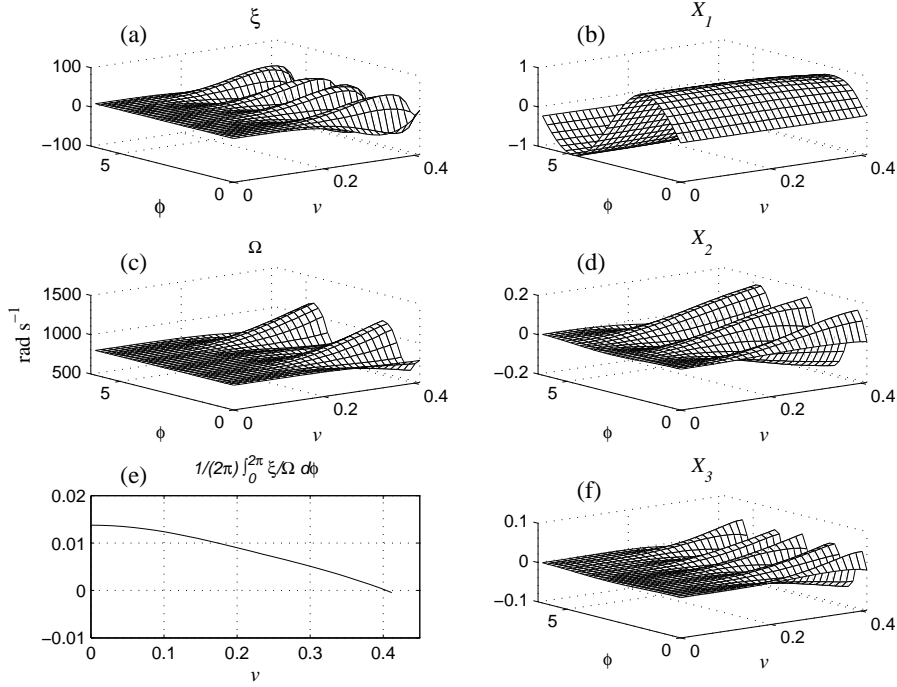} }
\caption{First nonlinear mode of the clarinet: (a) Damping function $\xi$, (b), (d) and (f) Surfaces for the three components of $\mathbf{X}$,  (c) Instantaneous frequency $\Omega$, (e) The scalar function $<\tau>$, the zero of which indicates an estimate of the amplitude of the limit cycle. \label{manifolds1}}
\end{figure}

In a first step, the NM described by $X$ and $Y$, as well as $\xi$ and $\Omega$, is computed solving
Eqs. (\ref{eq4a})--(\ref{eq7}) with the method described  section~\ref{sect-numsol} and
by choosing the first linear mode of the linearized system (which differs from the first linear mode of the
resonator) as the initial condition at $v=0$. The shapes of the surfaces $X$, $\xi$ and $\Omega$ are shown in Fig.~\ref{manifolds1} for a case with 31 discretization points in the $\phi$-direction (effectively resolving the first 15 Fourier terms), and 50 discretization points on the interval $v\in[0,0.41]$. As expected, the shape of $X_1$ starts out from a purely harmonic variation with respect to $\phi$ at $v=0$. The shape then changes only slightly as the amplitude grows. Components $X_2$ through $X_3$ are small at $v=0$, but then change dramatically as the amplitude grows. For low amplitudes, $\xi$ is constant and positive, which is characteristic for an unstable motion. As the amplitude grows, the shape becomes more complicated attaining both positive and negative values.
The function $<\tau>$ (Eq.~(\ref{eq1ddl16mm})) starts out at a positive value, and crosses the $v-$axis at
$v^{**}=0.4046$
 with a negative derivative.  This is a necessary condition for a self-sustained oscillation (see  section~\ref{NLMprop}).

\begin{figure}
\centerline{\includegraphics[width=8.5cm]{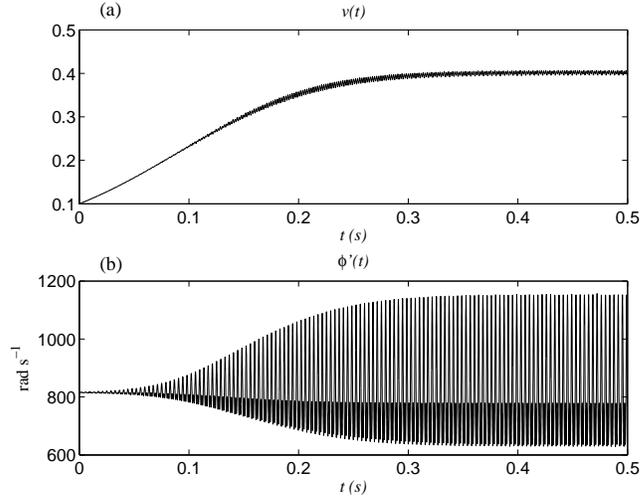} }
\caption{First nonlinear mode: (a) Time evolution of $v$, (b) Time evolution of $\dot{\phi}$. Barely discernible is a fine ripple in $v$ with the same period as the limit cycle. \label{vphi1}}
\end{figure}

\begin{figure}
\centerline{\includegraphics[width=0.9\columnwidth]{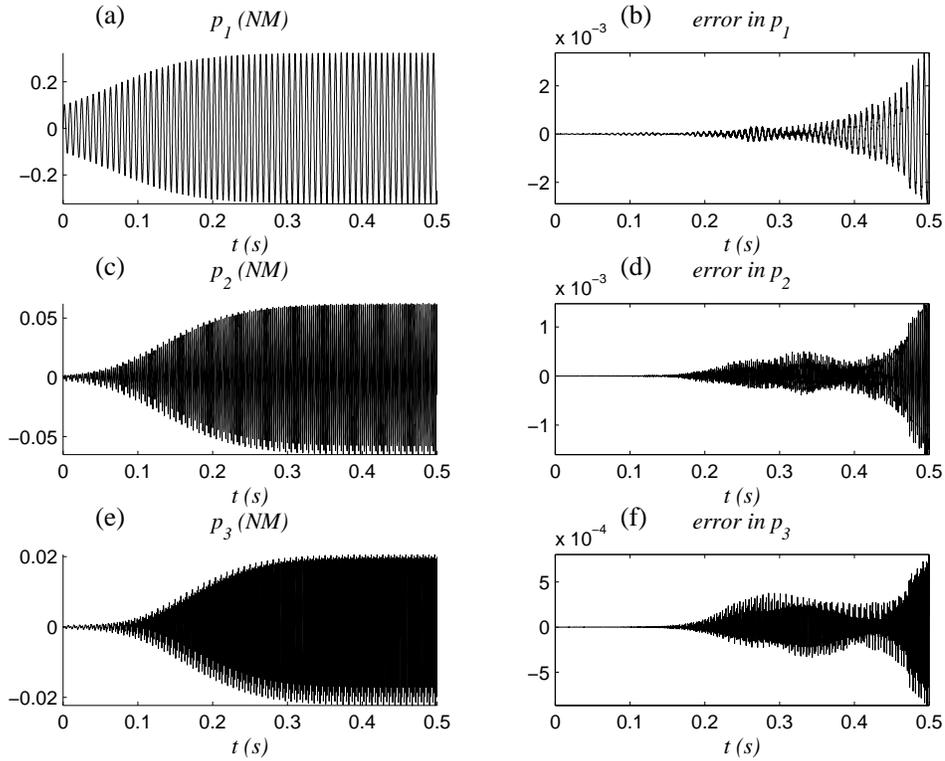} }
\caption{First nonlinear mode: (a), (c) and (e) Time evolution of $p_1$, $p_2$ and $p_3$ characterizing a modal motion according to the NM approach. (b), (d) and (f) Errors computed with a direct simulation of Eq.~(\ref{e:complete_model}) and the same initial conditions as reference.\label{psol1}}
\end{figure}

In a second step, a time evolution $v(t)$ and $\phi(t)$ is calculated from an arbitrary, small-amplitude initial condition ($v(0)=0.1$, $\phi(0)=0$) numerically solving Eq.~(\ref{eq3}), as described in section \ref{ss:NM-compute}. According to the obtained results, the time evolutions of the components of $\mathbf{P}(t)$ and $\dot{\mathbf{P}}(t)$ can finally be computed with Eq.~(\ref{eq2}). These results are shown in Fig.~\ref{vphi1}--\ref{psol1}. Instead of $\phi(t)$, the instantaneous frequency $\dot{\phi}(t)$ is shown.
As expected, for small values of $v$, $\dot{\phi}$ oscillates around  $815.55$ rad s$^{-1}$ which is the value of the imaginary part of the first eigenvalue of the linearized system.
It is noticeable  that there is a slight difference between this frequency and the corresponding resonance frequency of the $1/4-$wavelength resonator ($2\pi c/4l = 815.37$ rad s$^{-1}$). For larger values of $v$, the amplitude of the oscillation in $\dot{\phi}$ is increasing.
As can be seen in Fig.~\ref{psol1}, the amplitude of the first component $p_1$ grows quickly initially, but then stabilizes as the limit cycle is approached. The same phenomenon is visible for the higher components, but they show a much stronger relative growth. This is a typical feature for wind instruments, where the small amplitude oscillations are nearly sinusoidal. As the amplitude grows, nonlinear effects add increasingly to the timbre by successive enrichment of the harmonic content of the signal. The envelopes of the components $p_n$ are defined by the form of $v$ together with the evolution of the components of $\mathbf{X}$ versus $v$.

In order to check the validity of the solutions, a reference solution (with initial conditions  given by  Eq.~(\ref{eq2}) at $t=0$) was computed by direct solving of Eq.~(\ref{e:complete_model}) using a Runge-Kutta-Fehlberg solver (\verb/ode45/ in \verb/Matlab/) with a small tolerance. The difference between the solution obtained from the MN approach and the reference solution is presented as the error in Fig.~\ref{psol1}. The error is very small during the transient phase, but then starts to grow slowly in time mainly due to the error in the frequency. A longer simulation demonstrates (not shown in figure) that the error envelope grows for some time approximately linearly with time, consistent with a slowly growing phase lag. (The same kind of error growth would be prevalent for any method with numerical dispersion, however small the error in the frequency, albeit at a different rate.)

\begin{figure}
\centerline{\includegraphics[width=8.5cm]{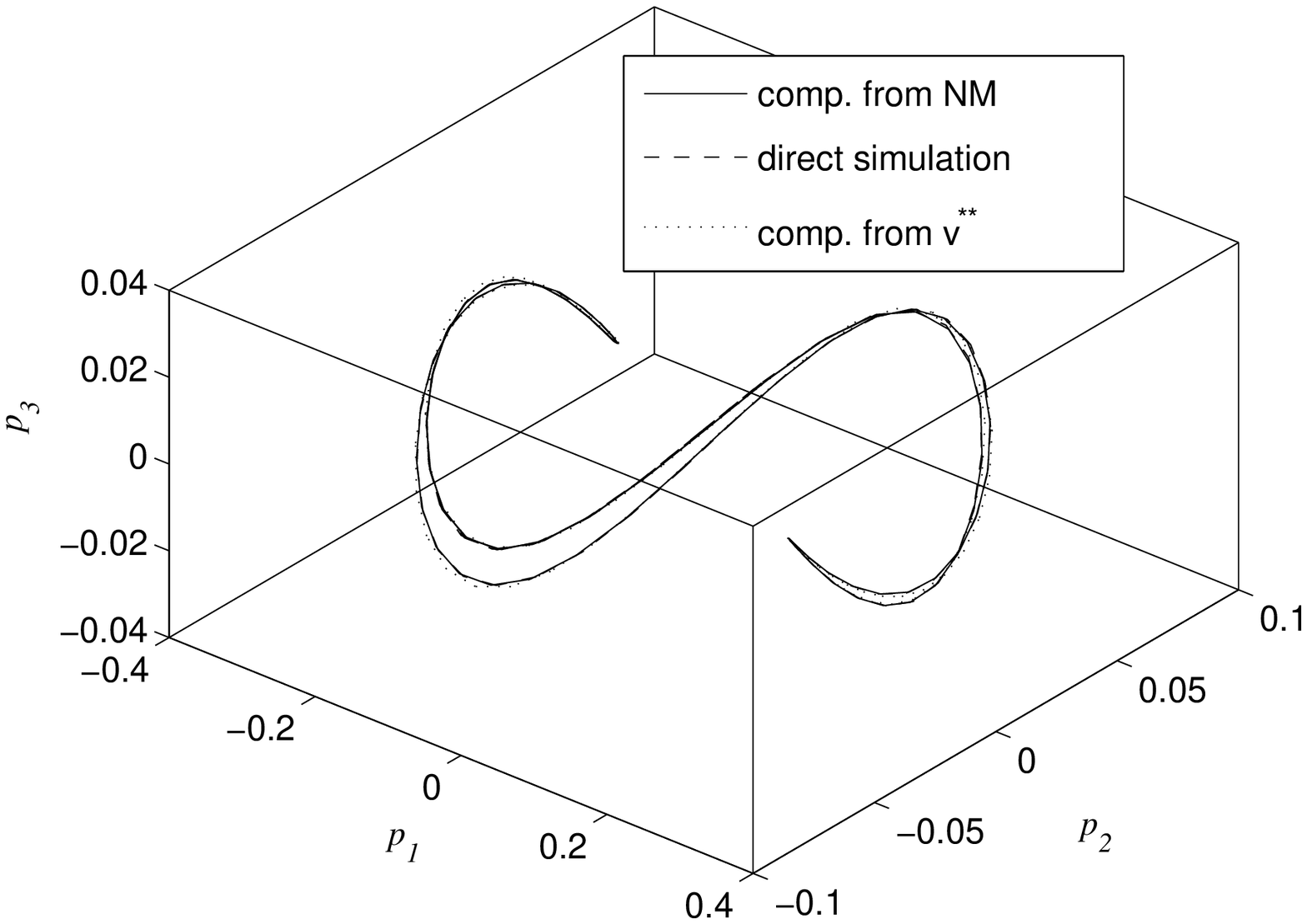} }
\caption{First nonlinear mode: The limit cycle in the configuration space $(p_1,p_2,p_3)$. 
--- Limit cycle computed from the nonlinear mode, - - - limit cycle according to direct simulation, ...... limit cycle computed from $v^{**}$. \label{fig:limitcyclemode1}}
\end{figure}

It is interesting to compare the limit cycle computed from the NM, with that obtained from the reference solution. The comparison eliminates any accumulated phase errors, and gives another more direct estimate of the error. Fig.~\ref{fig:limitcyclemode1} shows the limit cycles superimposed. The difference is hardly distinguishable in the plot. An approximation of the limit cycle computed from Eq.~(\ref{eq1ddl16bbb}), with the constant value $v^{**}=0.4046$ (zero of $<\tau>$, see bottom left of Fig.~\ref{manifolds1}), is also shown in the figure. Evidently, a good approximation of the limit cycle can also be obtained.
In order to get an estimate $A_p$ of the amplitude of the pressure signal $p=\Sigma_{i=1}^3 p_i$ in the steady state, surfaces $X_i$ must be taken into account through Eq.~(\ref{eq1ddl16bbb})
\begin{equation}
A_p =  v^{\ast\ast} \left(
\max_{\phi} \displaystyle \Sigma_{i=1}^3 X_i (v^{\ast\ast},\phi)
- \min_{\phi} \displaystyle \Sigma_{i=1}^3 X_i (v^{\ast\ast},\phi)
\right).
\end{equation}
The limit cycle frequency given by Eq.~(\ref{eq1ddl16bbbb}) is $815.2$ rad s$^{-1}$, which can be compared to the linear resonance frequency of the 1/4 wave resonator that is $815.37$ rad s$^{-1}$, and to the eigenfrequency of the linearized system that is $815.55$ rad s$^{-1}$.
Thus a modal motion according to the first NM corresponds to a clarinet sounding in its first register.
We have seen that the instantaneous amplitude and frequency are oscillating, even in the steady state. This is not in contradiction to the fact that the corresponding regime in terms of time evolution of $p_i$ is periodic. This can be checked from Fig.~\ref{fig:limitcyclemode1} where the dynamics of the steady state corresponds to a limit cycle in the configuration space.

\subsection{Second mode}

\begin{figure}
\centerline{\includegraphics[width=0.9\columnwidth]{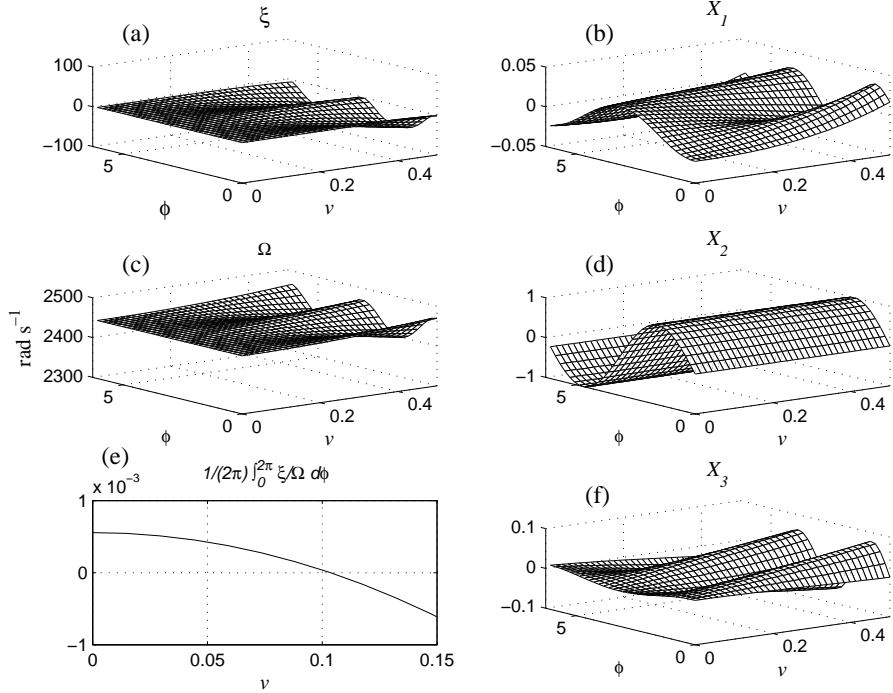} }
\caption{Second nonlinear mode of the clarinet: (a) Damping function $\xi$, (b), (d) and (f) Surfaces for the three components of $\mathbf{X}$,  (c) Instantaneous frequency $\Omega$, (e) The scalar function $<\tau>$. \label{manifolds2}}
\end{figure}

\begin{figure}\centerline{\includegraphics[width=8.5cm]{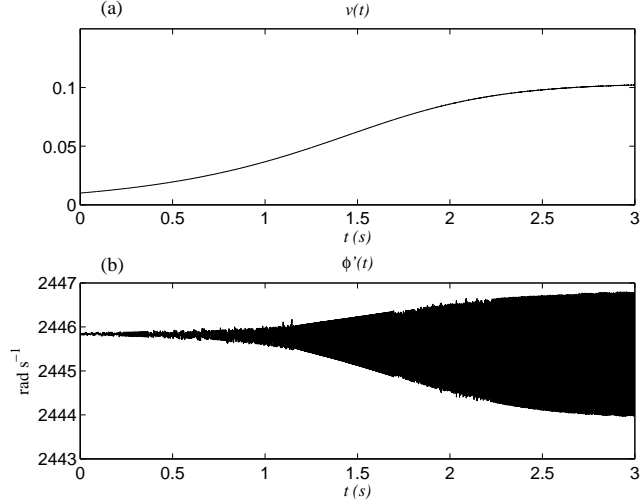} }
\caption{Second nonlinear mode: (a) Time evolution of $v$, (b) Time evolution of $\dot{\phi}$. \label{vphi2}}
\end{figure}

\begin{figure}
\centerline{\includegraphics[width=8.5cm]{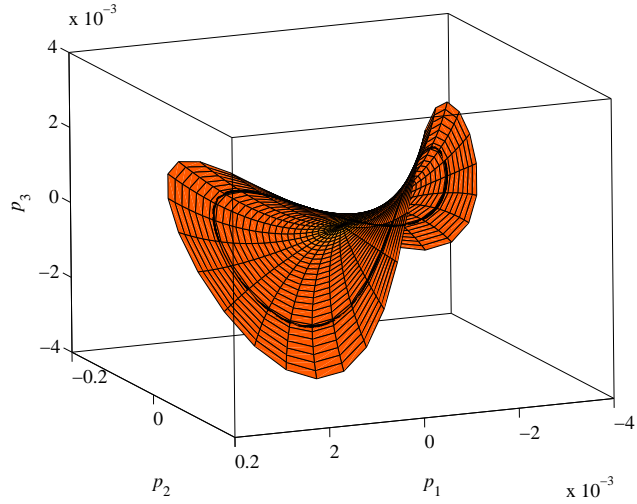} }
\caption{Second nonlinear mode: The limit cycle in the configuration space $(p_1,p_2,p_3)$ lying on the invariant manifold. Computations are carried out from the NM approach.\label{fig:manifoldcurve}}
\end{figure}

\begin{figure}
\centerline{\includegraphics[width=0.9\columnwidth]{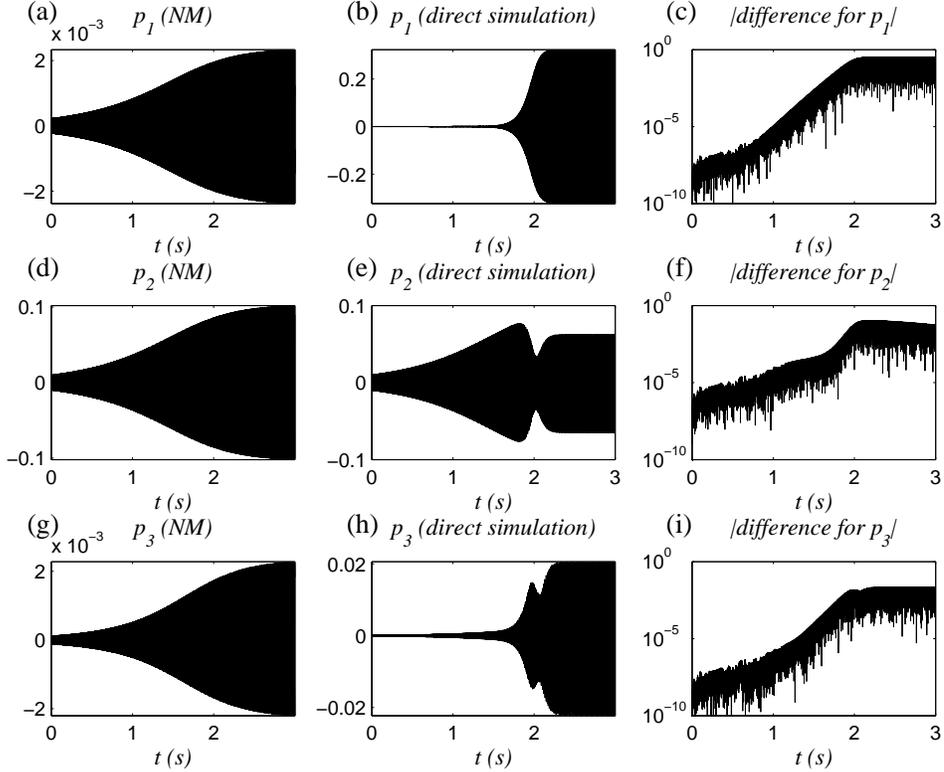} }
\caption{Second nonlinear mode: (a), (d) and (g): Time evolution of $p_1$, $p_2$ and $p_3$ characterizing a modal motion computed from the NM approach. (b), (e) and (h) Time evolution computed by direct simulation of Eq.~(\ref{e:complete_model}) using the same initial conditions. (c), (f) and (i) Differences between solutions computed in the two different ways.
 \label{psol2}}
\end{figure}

A similar investigation for the second nonlinear mode of the clarinet is presented in Fig.~\ref{manifolds2}--\ref{psol2}. The initial conditions for the computation of the nonlinear mode have been changed to the second linear mode of the linearized system.

We see in Fig.~\ref{manifolds2}  that the second component $X_2$ is now dominating over $X_1$ and $X_3$. The value of $\xi$ shows a variation that initially resembles that of the first mode, with positive as well as negative values. The amplitude $v^{**}=0.1035$ of the limit cycle is again found as the zero of $<\tau>$, which is smaller than for the first mode.
Also for the second mode, the derivative of $<\tau>$ is negative at $v=v^{**}$, showing that the limit cycle is stable \emph{on the invariant manifold}. This means that the limit cycle is stable with respect to a subspace of disturbances, but not necessarily to any disturbance.

Fig.~\ref{vphi2} shows the time evolution of $v$ and $\dot{\phi}$ calculated from an arbitrary, small-amplitude initial condition ($v(0)=0.01$, $\phi(0)=0$) solving numerically Eq.~(\ref{eq3}). It is noticeable that ripples on $v$ and $\dot{\phi}$ are much weaker than in Fig.~\ref{vphi1}.  This is related to the smoothness of the surfaces
$\xi$ and $\Omega$ (see Fig.~\ref{manifolds2}), which  remain much more regular than for the first mode,  even past the limit cycle amplitude.


Fig.~\ref{fig:manifoldcurve} shows the limit cycle of NM 2 on the invariant manifold represented by the three components $p_1$--$p_3$ according to Eq.~(\ref{eq1ddl16bbb}). The smoothness of the invariant manifold is linked with the smoothness of the surfaces of $\mathbf{X}$, which are again much smoother even for high amplitudes, than is the case for the first mode.
The invariant manifold for the first mode would be much more difficult to visualize due to its intricate folds and intersections with itself. The shown surface is really a projection of a six-dimensional manifold (including also $\dot{p}_1$--$\dot{p}_3$) that does not intersect with itself.

A direct numerical simulation (solving Eq.~(\ref{e:complete_model})) indicates that the limit cycle for the second nonlinear mode may be unstable in the phase space for the chosen parameter values. Indeed, a direct numerical simulation started from initial conditions on the manifold representing the second nonlinear mode, will eventually jump out of the manifold due to round-off and truncation errors. This is seen from the curves in the center column of Fig.~\ref{psol2}. The time evolution is initially the same as for the solution computed from the NM approach (left column), but after about 1.5 s, the first component starts to become notably excited and the growth of the second component is discontinued. The limit cycle for the second nonlinear mode is never reached, and the oscillation eventually converges to the limit cycle of the first mode.

To analyze the stability of the limit cycle, Floquet Theory has to be applied \cite{hale69}. Starting from the
approximated limit cycle given by Eq.~(\ref{eq1ddl16bbb}), the monodromy matrix is computed solving
the $2\pi$-periodic variational linear differential system associated to Eq.~(\ref{e:complete_model})
over one period, using the six canonical basis vectors as initial
conditions successively (see details in \cite{bb2006}). The computations show that the periodic orbit (approximated by Eq.~(\ref{eq1ddl16bbb}) ) associated with
the second NM is  unstable in the phase space (two complex conjugate
multipliers are outside the unit circle).

Thus, for this clarinet model, the second register appears unstable with the chosen parameter values, which is often experienced by beginners on real instruments.

\subsection{Third mode}

\begin{figure}
\centerline{\includegraphics[width=0.9\columnwidth]{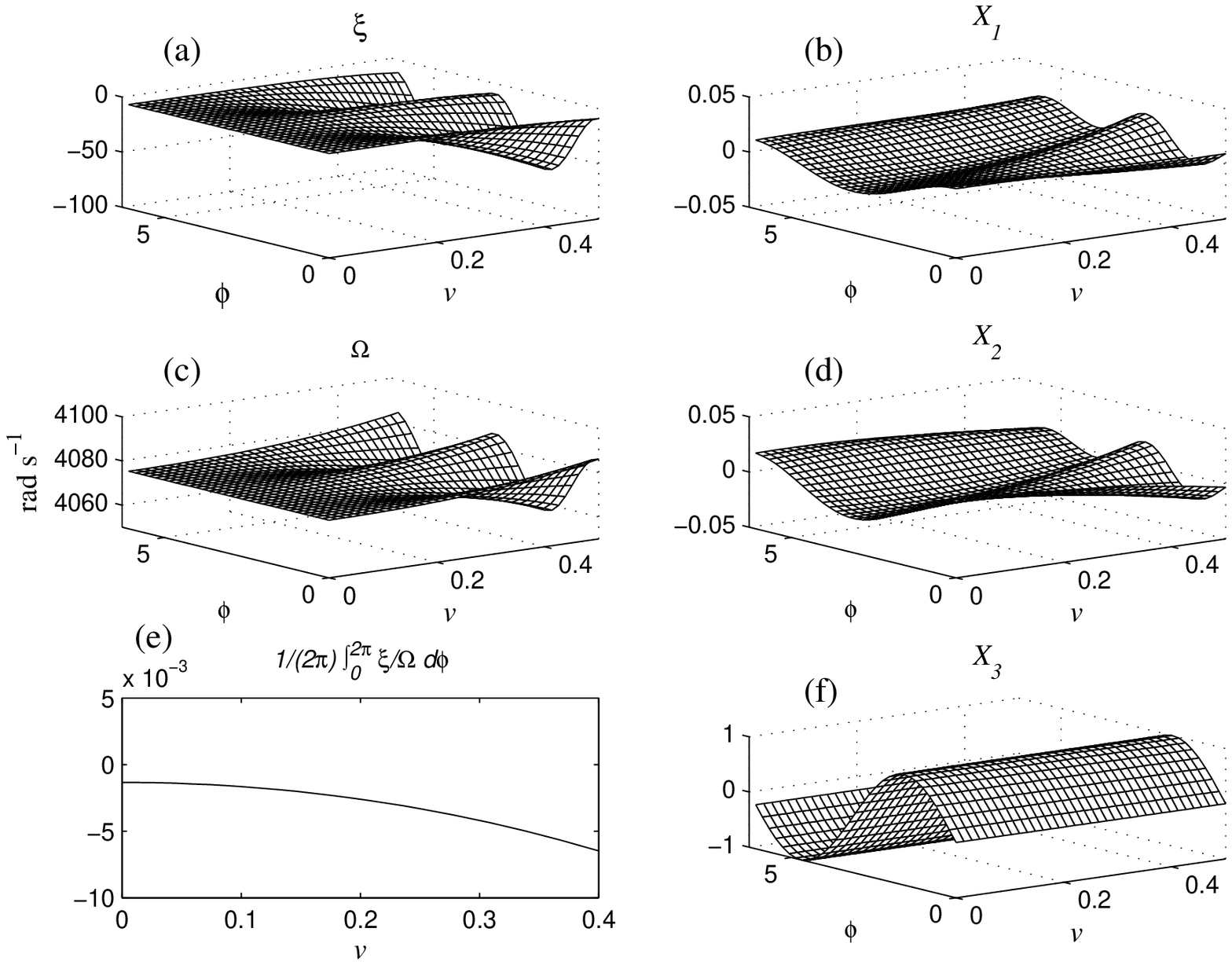} }
\caption{Third nonlinear mode of the clarinet: (a) Damping function $\xi$, (b), (d) and (f) Surfaces for the three components of $\mathbf{X}$,  (c) Instantaneous frequency $\Omega$, (e) The scalar function $<\tau>$. \label{manifolds3}}
\end{figure}

\begin{figure}
\centerline{\includegraphics[width=8.5cm]{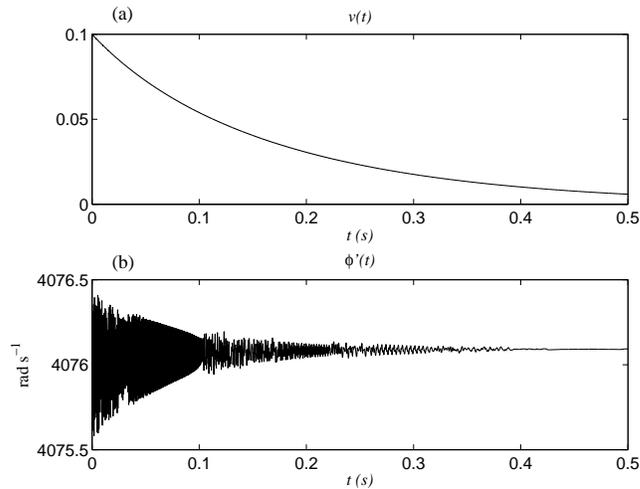} }
\caption{Third nonlinear mode: (a) Time evolution of $v$, (b) Time evolution of $\dot{\phi}$. \label{vphi3}}
\end{figure}

Calculations for the third NM are presented in Fig.~\ref{manifolds3}--\ref{vphi3}. The initial conditions for the computation of the NM have been changed to the third linear mode of the linearized system.

We see, Fig.~\ref{manifolds3},  that the third component $X_3$ is now dominating over $X_1$ and $X_2$.
The surface $\xi$, unlike for the first and the second modes, starts out at a negative value for $v=0$. As $v$ increases, the surface becomes increasingly oscillatory, but at no point is the mean value positive. In terms of the mean damping function, $<\tau>$ is strictly negative.
As a consequence, for the chosen blowing parameter $\gamma$, a solution started at any point on the third mode invariant manifold converges to the equilibrium point which is stable. There is no limit cycle in the invariant manifold associated to the third NM.
This is exemplified, in the time domain on $v$ and $\dot{\phi}$ with small-amplitude initial condition $v(0)=0.1$ and $\phi(0)=0$ (see Fig.~\ref{vphi3}).

\subsection{Discussion}

Some problems with divergence of the solution in the computation of the NM have been observed for certain values of $v$. Numerical experiments indicate that the value $v_{\mathrm{div}}$, where the instability occurs, converges to a certain value as the step size $h_v$ in the $v$-direction decreases. In the case of a model with one single degree of freedom, it is possible to formulate a version of Eqs.~(\ref{eq4a})--(\ref{eq4b}) where the algebraic constraints Eqs. (\ref{eq6})--(\ref{eq7}) are eliminated by assuming $\mathbf{X}(v(t),\phi(t))=\cos(\phi(t))$ in Eq.~(\ref{eq2}). The divergence persists also in that case, which indicates that the problem is not a consequence of the differential-algebraic structure of the original problem.
The findings suggest some sort of ill-posedness in the continuous equation, rather than a numerical problem, but the exact nature of this is yet to be investigated in detail.
For higher blowing levels, the point of instability is reached before the amplitude of stable oscillation is reached, thus limiting the applicability to medium amplitudes. However, the validity of the clarinet model itself is limited
to medium amplitudes (i.e. non beating reed).

The clarinet model with up to 8 degrees of freedom has been investigated, with results that are in accordance with the case $N=3$. There is in principle no limit of $N$, but the computational time increases with the size of the system.

\section{Conclusion}
We have seen that the suggested method for computing NM motions of dynamical systems, using amplitude and phase as master variables, is capable of dealing with self oscillating systems with internal resonances.
The numerical method presented in this paper offers a flexible way to handle the problem of numerical refinement for increased accuracy, and allows for the treatment of larger systems with many degrees of freedom -- unlimited in principle, but limited by memory demands and execution time considerations in practice. A bottleneck in the computations is the solution of Eqs.~(\ref{eq4a})--(\ref{eq7}). The implicit method requires the solution of a nonlinear system of equations, whose size grows with $N$ and $N_\phi$, but the use of an analytical Jacobian matrix speeds up the computation.
Although the calculation of the manifolds is unwieldy, it is a pre-processing step after which the computational work is independent of, or grows only linearly (the work of forming the components $p_j$) with $N$.
The surfaces $\mathbf{X}$, $\mathbf{Y}$, $\mathbf{\xi}$ and $\mathbf{\Omega}$ are intricate for amplitudes in excess of the limit cycle amplitude $v^{**}$, but for smaller $v$, their smoothness allows for a more compact reduced order representation, e.g. each surface might be represented by one single, or a piece-wise patchwork, of approximating functions with a limited number of parameters. This could greatly reduce storage demands and the need for interpolation.

This approach has many valuable by-products. For example, after having solved Eq.~(\ref{eq3}), which is an ODE with two unknowns, one has immediate access to the instantaneous amplitude and frequency without the need to solve the whole system (\ref{e:complete_model}). Moreover, one has also direct access to the solution components $p_1,\ldots,p_N$ and $\dot{p}_1,\ldots,\dot{p}_N$ for any $t$. We have also seen that is possible to compute unstable dynamics, including transients and the steady state. The limit cycles, finally, can be estimated without any computation whatsoever in the time domain.  However, the manifold (and the dynamics) are calculated for constant parameters $\gamma$ and $\zeta$. Therefore, a consequence highlighted by this study is that a direct application of the
approach for sound synthesis is therefore not obvious. On the contrary, the approach is predominately adapted for analyzing model dynamics.

\section*{Acknowledgments}
The authors want to thank Jean Kergomard for fruitful discussions.

The study presented in this paper was lead with the support of the French National Research Agency ANR winthin the CONSONNES project. The first author was financed by the French Ministry of Research through a post-doctoral grant.



\end{document}